# NetMoST: A network-based machine learning approach for subtyping schizophrenia using polygenic SNP allele biomarkers

*Xinru Wei*[1,2], *Shuai Dong*[2], *Zhao Su*[2], *Lili Tang*[1,5], *Pengfei Zhao*[1,5], *Chunyu Pan*[3], *Fei Wang*[1,5*], *Yanqing Tang*[6,7,8*], *Weixiong Zhang*[4*]，*Xizhe Zhang*[2*]

1. Early Intervention Unit, Department of Psychiatry, The Affiliated Brain Hospital of Nanjing Medical University, Nanjing, Jiangsu 210029, China;

2. School of Biomedical Engineering and Informatics, Nanjing Medical University, Nanjing, Jiangsu 210001, China

3. School of Computer Science and Engineering, Northeastern University, Shenyang, China

4. Department of Health Technology and Informatics, Department of Computing, The Hong Kong Polytechnic University, Hong Kong, China

5. Functional Brain Imaging Institute of Nanjing Medical University, Nanjing, China

6. Department of Psychiatry, The First Affiliated Hospital of China Medical University, Shenyang, China

7. Brain Function Research Section, The First Affiliated Hospital of China Medical University, Shenyang, China

8. Department of Gerontology, The First Affiliated Hospital of China Medical University, Shenyang, China

*: Correspondence: FW: fei.wang@yale.edu, YT: yanqingtang@163.com, WZ: weixiong.zhang@polyu.edu.hk, XZ: zhangxizhe@njmu.edu.cn

**Abstract**

Subtyping neuropsychiatric disorders like schizophrenia is essential for improving the diagnosis and treatment of complex diseases. Subtyping schizophrenia is challenging because it is polygenic and genetically heterogeneous, rendering the standard symptom-based diagnosis often unreliable and unrepeatable. We developed a novel network-based machine-learning approach, netMoST, to subtyping psychiatric disorders. NetMoST identifies polygenic risk SNP-allele modules from genome-wide genotyping data as polygenic haplotype biomarkers (PHBs) for disease subtyping. We applied netMoST to subtype a cohort of schizophrenia subjects into three distinct biotypes with differentiable genetic, neuroimaging and functional characteristics. The PHBs of the first biotype (36.9% of all patients) were related to neurodevelopment and cognition, the PHBs of the second biotype (28.4%) were enriched for neuroimmune functions, and the PHBs of the third biotype (34.7%) were associated with the transport of calcium ions and neurotransmitters. Neuroimaging patterns provided additional support to the new biotypes, with unique regional homogeneity (ReHo) patterns observed in the brains of each biotype compared with healthy controls. Our findings demonstrated netMoST's capability for uncovering novel biotypes of complex diseases such as schizophrenia. The results also showed the power of exploring polygenic allelic patterns that transcend the conventional GWAS approaches.

**INTRODUCTION**

Schizophrenia (SCZ) is a devastating neuropsychiatric disorder with considerable morbidity and mortality[1-3] and exerts substantial health and socioeconomic burdens[4]. It has a strong genetic predisposition and heritability estimated at 60-80%[5,6]. Significant heterogeneities in etiology, pathophysiology and symptom characterize the disease. SCZ shares genetic and clinical characteristics with other psychiatric disorders, such as bipolar disorder (BD) and major depressive disorder (MDD)[7-9]. The current diagnostic criteria for SCZ are primarily based on behavioral and cognitive indicators[10-12], which are challenging to quantify, inconsistent and subjective. As a result, SCZ diagnosis often results in poorly informed therapeutic strategies, unpredictable treatment outcomes, and frequent relapses.

To explore and exploit the abovementioned heterogeneities to improve the diagnosis and treatment of SCZ, it is crucial to adequately delineate SCZ into subtypes that provide deep insight into disease pathogenesis, pathophysiology, and therapeutic options. One common approach for subtyping SCZ is based on the presence or absence of symptom profiles. For example, Dickison et al.[13] identified two subgroups of SCZ based on psychotic symptoms, a "deficit" subtype and a "distress" subtype. Lim et.al[14] discovered four homogeneous groups of individuals with SCZ of different severity of cognitive impairment. However, SCZ symptoms may vary depending on the illness stages, comorbid conditions, and medication[15]. This instability can make it challenging to classify patients into proper subtypes and make it difficult to track disease progression or predict treatment response. Moreover, symptom-based subtyping provides little information on the underlying disease mechanisms and translates poorly into effective therapeutic strategies. Another subtyping approach is based on objective neurobiological endophenotypic features, including MRI-based neuroanatomical measures[16,17], functional magnetic resonance imaging (fMRI)[18], and combinations of electrophysiology and cognition[19]. Despite extensive research, valid neurobiological subtypes of SCZ remain elusive. This is primarily because the number of subtypes and their underlying biological features vary across studies, even when utilizing the same neuroimaging features. For example, Chand et al.[16] and Xiao et al.[17] independently identified two and three distinct neuroanatomical SCZ subtypes using gray matter volume as a neurobiological marker.

Manifestation of psychiatric symptoms and endophenotypes, such as brain activities and functional connectivities, result from the interaction between genetic predispositions and environmental triggers. Therefore, genetic risk factors are objective and more reliable biomarkers for inheritable psychiatric disorders like SCZ. However, little has been done to leverage SCZ genetics for diagnosis. Luo et al.[20] used a peripheral epigenome-wide DNA methylation array to cluster 63 SCZ patients into two biotypes. One biotype displayed prominent methylation abnormalities and was associated with dysregulated immune function. Another recent study[21] applied multi-view clustering to clinical and single nucleotide polymorphism (SNP) data and grouped a small group of SCZ subjects into three biotypes with differential cognitive measures and disease courses. While these methods used genotyping data, they did not identify subtype-specific genetic risk factors that might be used as potential biomarkers for the disease.

Subtyping and identifying genetic biomarkers for SCZ are difficult because the two are intertwined. Discovering biomarkers for complex diseases typically requires comparing diseased and healthy groups to identify critical genetic features. However, individuals with a complex disease are often genetically heterogeneous and may carry multiple distinct biomarkers. This genetic heterogeneity can impede our ability to identify statistically significant biomarkers from the entire samples, leading to inconsistent and non-reproducible results. Overcoming this impediment requires subtyping the subjects and identifying biomarkers for every subtype. On the other hand, disease subtyping involves partitioning samples into groups with homogenous members. The similarity among group members depends on selecting the right patient features. Therefore, it is necessary to find essential biomarkers as features to quantify similarity for subtyping. However, as discussed earlier, biomarker identification depends on accurate subtyping.

Therefore, disease subtyping and biomarker identification are interrelated. To our knowledge, no method has been developed for this issue.

Another challenge in subtyping SCZ is its polygenicity. Complex diseases, including SCZ, are caused by the interactions or associations of multiple genetic factors. Genome-wide association studies (GWAS) can help identify individual candidate genetic risk factors. However, these studies have limited power in finding genetic risk factors for complex diseases[22,23]. GWAS methods are designed to look for individual risk factors, so they fail to identify groups of related genetic factors, each of which may have small effect sizes when examined in isolation[24]. Several approaches, such as the polygenic risk score (PRS)[25], have been attempted to address these limitations by looking for combinations of genetic variants to increase statistical power. However, these methods are designed for evaluation rather than discovery. Therefore, "present PRSs typically explain only a small fraction of trait variance"[26].

We propose a novel _network module-based method for subtyping_ (netMoST) to simultaneously address the challenge of subtyping and identifying polygenic biomarkers for complex diseases. Taking SCZ as a case study, we applied netMoST to group SCZ into three biotypes. The new biotypes were validated utilizing multimodal data, including genetic variations, brain neuroimaging patterns and clinical features. We identified subtype-specific polygenic risk factors to understand the pathogenetic and pathophysiological features of SCZ. A noteworthy finding is an SCZ subtype defined by modules of genomic alleles on immune genes in the major histocompatibility complex of the human genome, suggesting the involvement of immunity in the pathogenesis and pathophysiology of SCZ.

**RESULTS**

*Identifying subtypes and polygenic haplotype biomarkers for complex disease*

Complex diseases, such as schizophrenia (SCZ), are polygenic, meaning that variations at multiple genomic loci rather than individual ones drive the onset of disease phenotypes and disease progression. Many complex diseases are also genetically heterogeneous, meaning that the same symptoms in different patients may be attributed to various genetic factors, indicating the existence of multiple disease mechanisms. As a result, complex diseases like SCZ have numerous polygenic risk factors. Genetic heterogeneity and multiple disease mechanisms make it necessary to adequately define diseased cases into biotypes for accurate diagnosis and effective treatment.

We developed a novel _network module-based method for subtyping_ (netMoST) complex diseases like SCZ (Figure 1). Our new approach hinged upon two ideas: focusing on SNP alleles instead of SNPs to deal with genetic heterogeneity and adopting network analysis to facilitate the identification of polygenic biomarkers. The new algorithm first constructs an SNP-allele network based on the Custom Correlation Coefficient (CCC), a measure designed to capture genetic heterogeneity by computing a multi-faceted collection of correlation[27,28]. The algorithm then detects communities in the network to identify modules of functionally associated SNP alleles. Modules of multiple SNP alleles that are statistically significantly correlated with the disease are selected as polygenic features. Unsupervised clustering is then applied to group the disease into biologically homogeneous subgroups. The subtypes are further characterized by subtype-specific polygenic SNP-allele modules using the same SNP-allele module finding scheme.

The central theme of netMoST is defining and identifying polygenic risk factors. We conceptualize polygenic risk factors as modules of highly correlated SNP alleles across the entire genome. Utilizing SNP alleles can address genetic heterogeneity[28,29], and SNP-allele modules may capture long-range genomic associations within a chromosome and across different chromosomes. Moreover, SNP-allele modules are specific to SCZ subjects, enabling distinction between SCZ cases and healthy controls. We termed these SNP-allele modules *polygenic haplotype biomarkers* (PHBs).

We formulated the problem of finding PHBs as a problem of identifying modules of highly correlated nodes in a network of SNP alleles. We split the problem into two. The first was constructing a network of nodes representing SNP alleles and edges representing correlations between pairs of SNP alleles using the CCC measure[27]. The second problem was the general problem of finding modules of highly correlated nodes in a network, which has been extensively studied[30]. We used the Louvain network-module detection method[31] in the current implementation of netMoST.

### NetMoST defines three biological subtypes of schizophrenia

*Polygenic haplotype biomarkers define three biotypes of schizophrenia*

We applied NetMoST to subtype schizophrenia. A total of 425 participants were recruited, including 141 SCZ patients and 284 healthy controls (HCs). After preprocessing and quality control of the genotyping data (see Methods), we identified 404,078 SNPs for the 141 SCZ patients and 283 HCs (Figure 2A). These SNPs were used to construct an SNP-allele network where 480,339 interactions connected 192,409 nodes (SNP alleles). The Louvain network-module finding method[31] detected 36,576 SNP-allele modules from the network. Among these modules, 426 had Odds Ratios (ORs) and Risk ratios (RRs) greater than 1.2 (Figure 2A) and were taken as SCZ risk SNP-allele modules for subtyping the disease. The 426 risk modules spanned over 8,953 (4.7% of 192,409) SNP alleles, had an average of 19.4 alleles and a maximum of 150 alleles per module and had ORs ranging from 1.23 to 8.22 with an average of 1.78 (Figure 2B, Supplemental File 1). When analyzed separately, the individual SNPs within the risk modules had smaller OR values distributed around 1, indicating that they contributed insignificantly to SCZ phenotypes individually (Figure 2B). In contrast, the OR values of the SNP-allele modules were statistically significant, showing that schizophrenia was a polygenic disorder and suggesting that the disease was due to the interaction among the variations in multiple genes or genomic loci.

Taking the 426 risk SNP-allele modules as features, we define three subtypes or clusters of the SCZ subjects using the Principle Component Analysis (PCA)[32] and the KMeans clustering algorithm[33]. Biotype 1 consisted of 36.9% (n = 52) of the SCZ participants, biotype 2 of 28.4% (n = 40), and biotype 3 of 34.7% (n = 49) (Figure 2C). Demographic and clinical details of the biotypes are in Supplementary eTables 1 and 2.

We applied the same network construction and module detection methods to every biotype to discover subtype-specific risk SNP-allele modules using more stringent criteria: OR ≥ 1.5, a lower limit of 1.1 for the 95% confidence interval (CI), and Yates's Correction p-value < 0.05. We adopted subtype-specific risk SNP-allele modules as PHBs of SCZ. We found 264 PHBs for biotype 1, 83 PHBs for biotype 2, and 144 PHBs for biotype 3. Notably, the ORs of PHBs of the three biotypes were significantly greater than before subtyping (Figure 2D), showing that subtyping can enhance the detection of genetic risks of SCZ. Among the three biotypes, biotype 1 had the smallest *p*-value (biotype 1: 3.55E-59, biotype 2: 2.27E-31, biotype 3: 3.20E-41) and the most significant mean OR (biotype 1: 2.93, biotype 2: 2.84, biotype3: 2.73), indicating that it was the most distinct from HCs among the three biotypes.

To gain insight into the unique etiologies of the three SCZ biotypes, we performed a gene function enrichment analysis on the host genes of or genes nearest to the SNPs in the PHBs of each biotype using the FUMA online portal[34] (Figures 2E-2G). The SNP alleles in the biotype-1 PHBs were primarily associated with genes related to neurodevelopmental processes, including neurogenesis and neuron development and differentiation (Figure 2E). These biological pathways suggest that biotype-1 PHBs significantly influence neural systems, making biotype 1 a representative of the neurodevelopment of SCZ. The biological functions of SNP alleles in the biotype-2 PHBs were enriched considerably with immune functions, including regulating immune processes, immune response, defense response, and innate immune response (Figure 2F). These findings highlighted the relationship between immunity or inflammation and SCZ, suggesting that biotype 2 may constitute a highly inflammatory SCZ subgroup whose genetic variations lead to immune dysfunction and inflammation and contribute to SCZ pathophysiology. The SNP-alleles in the

biotype-3 PHBs were enriched in the transport of calcium ions and neurotransmitters (Figure 2G), which are mediators of physiological functions in the central nervous system and have been implicated in the pathogenesis of psychiatric diseases. Some genetic risk factors of biotype 3 are also involved in glycoprotein biosynthetic processes, suggesting glycosylation's role in SCZ. These results revealed distinctive genetic patterns of the three biotypes, reflecting a significant heterogeneity among them.

*Subtype-specific polygenic haplotype biomarkers*

To assess the power of subtype-specific PHBs as biomarkers, we compared their effect sizes and the risk SNPs chosen by the standard association analysis using the logistic regression model in PLINK v1.9[35]. Every patient in biotype 1 possessed at least 158 (60%) of the 264 PHBs. Every individual in biotype 2 carried at least 49 (59%) of the 83 PHBs, with a mean of 61 PHBs for all SCZ subjects. In biotype 3, every subject was covered by at least 99 (69%) of the 144 PHBs, with an average of 109 (76%) PHBs across all patients. Therefore, these PHBs are robust biomarkers for SCZ. Approximately 31.5% of the PHBs (biotype 1: 39.8%, biotype 2: 21.7%, and biotype 3: 22.2%) contained SNP loci spanning more than one chromosome, many of which were located in noncoding and intergenic regions.

To characterize the PHBs as genetic signatures of the biotypes, we analyzed their potential biological functions in SCZ. We present the results on the PHB with a significant odds ratio (OR) and a sufficient number of SNP alleles in genic regions from every subtype. We analyzed the potential functions of these PHBs using the functions of the host genes of the SNP alleles in the PHBs. Notably, the top 10 ranked PHBs of biotype 1 had significant ORs, ranging from 4.22 to 57.68 (Supplemental File 1). Many SNP alleles in these PHBs were located in noncoding or intergenic regions. We selected a representative PHB with 84.5% of SNP-alleles in protein-coding genes for functional analysis. This PHB appeared in 30% of the biotype-1 cases but 7.2% of HCs, giving rise to an OR of 5.46 (CI: 2.56-11.65). It was composed of 71 SNP alleles from 11 genes (*GRM5*, *NOTCH4*, *SORBS2*, *IL1RAPL1*, *STK19*, *TNXB*, *PPT2*, *RNF5*, *PBX2*, *TMEM132D*, and *FLRT2*) spanning eight chromosomes (Figure 3A, eFigure 1), representing a long-range haplotype over multiple genes across multiple chromosomes (eFigure 1). In contrast, the standard association analysis using PLINK detected only one SNP (rs9994907 in gene *SORBS2*) with statistical significance (OR: 1.56, *p*-value < 0.05, logistic regression analysis) (Figure 3B). The SNP-alleles of this PHB in *NOTCH4* (Notch Receptor 4) on chromosome 6, *GRM5* (Glutamate Metabotropic Receptor 5) on chromosome 11, and *IL1RAPL1* (Interleukin 1 Receptor Accessory Protein Like 1) on chromosome X exhibited significant network centralities, including average neighbor degrees, closeness centralities, and pagerank indices (eFigure 2). These three genes were involved in the development of SCZ. Specifically, fifteen SNP alleles were in *GRM5*, which has been regarded as a promising target for treating cognitive deficits of SCZ[36]. *NOTCH4*, hosting eight SNP alleles in this PHB, was known to be strongly associated with SCZ[37]. *IL1RAPL1* resided in a critical region on chromosome X that has been reported to be associated with a non-syndromic form of X-linked intellectual disability[38]. *IL1RAPL1* was expressed abundantly in post-natal brain structures involved in the hippocampal memory system, suggesting a critical role in regulating physiological processes underlying memory and cognition abilities[38]. These results revealed that the SNP alleles in this PHB might function in neural system development and cognition.

The PHB for biotype 2 consisted of 71 SNP alleles and, surprisingly, were distributed across nine chromosomes (Figure 3C). Twenty-five percent of the SCZ subjects in biotype 2 carried this PHB, compared to 6.2% of healthy controls, resulting in the highest OR of 5.04 (CI: 2.03-12.50) for biotype 2. Moreover, 43 (60.6%) of the SNP alleles of this PHB resided in the major histocompatibility complex (MHC) on chromosome 6, which is critical to immunity[39] (eFigure 3). Two SCZ-susceptible genes, *HLA-DQA1* and *HLA-DQB1*[40,41], are also located in this region (Figure 3C, eFigure 3). Furthermore, we computed various centrality measures for every SNP allele in the PHB, including the degree centrality[42], average neighbor degree[43], closeness centrality[44], and pagerank index[45,46]. The result showed that the SNP alleles in the MHC region had greater centrality, average neighbor degree, closeness centrality, and pagerank index, indicating their functional importance (eFigure 4). We also compared the OR for every SNP in this PHB using the association analysis based on the logistic regression model in PLINK[35]. Fifteen (21.1% of the 71) SNPs were statistically significant ($p$-value < 0.05). The SNP with the highest OR (MHC:rs9268199 in gene *C6orf10*) appeared in 28.8% of the cases in biotype 2 versus 12.2% of the healthy controls, resulting in an OR of 2.96 (CI: 1.71- 5.11), significantly smaller than the OR of 5.04 for the top-ranked PHB (Figures 3C-3D).

Biotype 3 had a representative PHB with an OR of 12.39 (CI: 3.46-44.32). It appeared in 15.56% of the biotype-3 subjects but 1.47% of HCs. This PHB comprised 67 SNP alleles on thirteen chromosomes. Most SNP alleles existed in three genes, *CDKAL1*, *PTPRD*, and *CACNA2D1* (Figure 3E, eFigure 5). Twenty-six SNP alleles were located in *CDKAL1*, a gene that rendered individuals with SCZ predisposed to type 2 diabetes, suggesting CDKAL1 as a shared genetic risk factor for both diseases[47]. The SNP alleles in *CDKAL1* had a greater centrality and average neighbor degree, highlighting their importance in this PHB. Gene *PTPRD* contained eight SNP alleles, which encoded a molecule for signal transduction, may function as a crucial neuronal cell adhesion molecule and synaptic specifier[48]. *CACNA2D1*, hosting four SNP alleles, encoded a voltage-gated calcium channel-related gene and was critical for mediating intracellular $Ca2^+$ influx and responsible for signaling transmission across synapses[49]. Like the other two biotypes, the SNPs selected by the single-marker method PLINK had much smaller ORs, detecting only one statistically significant candidate SNP (Figure 3F).

In summary, the PHBs identified by netMoST for the three biotypes had more significant ORs than the single-SNP counterparts determined by standard association analysis using the logistic regression model in PLINK (Figure 3). The significance of this result was multifold. First, netMoST was able to capture genetic risks that were missed by the conventional single-marker-based methods such as PLINK. In particular, netMoST could identify combinatorial, polygenic risk factors spanning multiple chromosomes (Figures 3A, 3C and 3E), which none of the existing genome-wide association studies could detect. These genetic risk factors were large haplotypes representing long-range linkage associations across the genome (eFigures 1, 3, 5). Third, using SNP alleles instead of SNPs could stratify SCZ diagnoses with subtle genetic or behavior differences and reveal proper latent biotypes. Fourth and importantly, the PHBs identified by netMoST were excellent genetic signatures of SCZ with sufficient effect sizes, making them ideal biomarkers for SCZ diagnosis.

*Subtype-specific alterations of neuroimaging patterns*

We identified significantly differential alterations in brain neuroimaging patterns among the three biotypes. While the regional homogeneity (ReHo) was significantly elevated in various frontal lobe regions of the three biotypes compared to HCs, the three biotypes had distinct brain regions displaying decreased ReHo. Biotype 1 had considerably increased ReHo in the frontal lobe (inferior frontal gyrus, middle frontal gyrus, and orbital part of the inferior frontal gyrus) but significantly decreased ReHo in the occipital lobe, cuneus, and lingual gyrus compared to HCs (Figure 4A). Biotype 2 had considerably increased ReHo in the frontal lobe regions of the superior frontal gyrus, middle frontal gyrus, and medial frontal gyrus but significantly decreased ReHo in the occipital lobe and lingual gyrus compared to HCs (Figure 4B). Biotype 3 had significantly increased ReHo in the frontal lobe (superior frontal gyrus, medial of superior frontal gyrus, and orbital part of the inferior frontal gyrus) but significantly reduced ReHo in the parietal lobe, postcentral gyrus, precentral gyrus, and temporal lobe (Figure 4C). These distinct intermediate neuroimaging phenotypes in ReHo among the three biotypes support these biotypes' distinctive polygenic risk factors and illustrate the power of netMoST as a diagnostic tool.

*Subtype-specific clinical characteristics*

To further assess and gain additional insight into the new biotypes defined by netMoST, we differentiated behavioral symptoms among the biotypes using the Brief Psychiatric Rating Scale (BPRS) and one-way ANOVA analysis (see Methods). The three biotypes differed significantly in three of the five BPRS factors scores, including Anxiety and Depression, Hostility-suspicion, and Thinking Disorder (Figures 4D-4F). Biotype 2 exhibited more severe behavioral symptoms in Hostility-suspicion and Thinking Disorder. Specifically, biotype 2 had a significantly higher level of Hostility-suspicion than Biotype 1 (p-value of 0.002 after applying the Least Significant Difference (LSD) correction) (Figure 4D), and also had a higher level of Thinking Disorder than the other two biotypes (*p*-value of 0.007 after LSD correction) (Figure 4E). Biotype 3 showed more severe Anxiety and Depression, as indicated by the lowest Anxiety and Depression score (*p*-value of 0.04 after LSD correction) (Figure 4F).

Similarly, one-way ANOVA analysis showed differences in cognitive ability among the three biotypes and HCs. The analysis utilized two cognitive measures, the MATRICS Consensus Cognitive Battery (MCCB) and Wisconsin Card Sorting Test (WCST). The three biotypes exhibited varying levels of cognitive impairments compared to HCs, as evidenced by smaller T-scores of MCCB and more significant scores of non-perseverative error (NPE) on WCST (Figures 4G and 4H). Specifically, biotype 1 had the worst cognitive performance, as indicated by the smallest T3 score of MCCB and the largest NPE score of WCST. The T3 score of MCCB for biotype 1 was significantly smaller than that for the other two biotypes (*p*-values of 0.03 and 0.04 after LSD correction) (Figure 4G), and the NPE score of WCST was substantially greater than biotype 3 (*p*-value of 0.02 after LSD correction) (Figure 4H). All these results supported SCZ subtyping and the new biotypes defined by netMoST.

**DISCUSSION**

The current SCZ diagnosis is symptom-based without considering objective indicators such as genetic risk factors. The disease's complex polygenicity and high genetic heterogeneity imply that SCZ may be further grouped into subtypes, as suggested in previous studies[50,51]. Proper subtypes of SCZ can be used to improve diagnosis and treatment. However, reliable subtypes of SCZ remain elusive due to the lack of genetic biomarkers for this polygenic brain disorder. Even though previous GWAS identified more than one hundred individual disease-associated SNPs[6], identifying genetic biomarkers or risk factors for polygenic diseases like SCZ remains challenging, even with a large sample size[52]. Such individual SNPs provide little information on the orchestrated interactions and associations among multiple genetic elements and have minuscule effect sizes, so they cannot be used as genetic biomarkers.

We developed netMoST, a novel approach for finding polygenic biomarkers and disease subtyping designed to address the polygenicity and genetic heterogeneity of complex diseases. We aimed to identify polygenic biomarkers and define adequate biotypes of SCZ. We formulated discovering polygenic biomarkers as a problem of identifying module structures in large SNP-allele networks and disease subtyping as a problem of network-based clustering. NetMoST hinged upon two critical ideas. First, we adopted SNP alleles instead of SNPs as the basic unit of analysis so that genetic heterogeneity could be adequately deconvoluted. Combined with CCC[27], a correlation metric designed for SNP alleles, this SNP-allele representation could detect subtle genetic heterogeneity and identify the individual patients who contributed to this property. Second, we adopted a network perspective on polygenic risk factors (Figure 1). This novel network perspective was materialized by taking advantage of the latest advancement in network sciences[53-56], particularly network module structure analysis[57]. Indeed, netMoST could effectively extract SNP-allele modules, many of which were statistically significant and functionally relevant and can be used as disease-specific PHBs (Figures 3A-3F, Supplemental File 1). Remarkably, these PHBs were statistically more significant than individual SNPs detected by the conventional GWAS methods and had substantially larger effect sizes. They were excellent candidate disease biomarkers for subtyping (as done in the current study) and diagnosis (to be studied in future clinical trials). The SCZ subtyping results were well supported by data and information from additional sources, including the potential involvement of the identified SNPs in relevant biological processes, coherent neuropsychiatric brain imaging patterns, and cognitive measures (Figures 2E-2G, 4A-4H).

The novel netMoST subtyping method produced three biotypes with distinctive characteristics that provide new insight into SCZ disease mechanisms. Biotype 1 was a typical subtype of SCZ with more severe cognitive impairment and showed a neurodevelopment-associated genetic pattern, whose genetic variations were involved in neuron development and differentiation as neurogenesis (Figure 2E). The subtype-specific PHB of biotype 1 emerged as the pattern of interactions between multi-genes across different chromosomes (Figure 3A, eFigure 1). The crucial genetic risk factors, including a combination of *NOTCH4*, *GRM5*, and *IL1RAPL1*, were identified through network analysis (Figure 3A, eFigure 2). These genes have been indicated to affect memory and learning abilities, and brain development, each of which was reported in previous studies[36-38,58]. Indeed, this subtype may be caused by the interactions of genetic variations, representing neurodevelopmental abnormalities. Additionally, the ReHo of biotype 1 was significantly decreased in the frontal lobe and increased in the occipital lobe, cuneus, and lingual gyrus compared to HCs.

Significantly different from biotype 1, biotype 2 was primarily characterized by immune/inflammatory regulation involvement. The PHB with the highest OR showed a high correlation with immunity. Most of its genetic variations resided in the MHC region, which is known to harbor genetic variants conferring the risk of SCZ[59]. Network analysis also emphasized the importance of the MHC region in the subtype-specific PHB of biotype 2 (eFigure 4). Consistent with the previous study, the findings that two key genes hosting these alleles, *HLA-DQA1* and *HLA-DQB1*, were identified and further demonstrated the association of this biotype with the immune system. More evidence exists for the involvement of immune proteins in the function of the central nervous systems, such as differentiation and migration and synaptic plasticity[60,61]. These core biological processes are related to SCZ pathogenesis[62]. Furthermore, biotype 2 had significantly increased ReHo in the frontal lobe and significantly decreased ReHo in the occipital lobe and lingual gyrus compared to HCs. Conceivably, immune-related genetic variations may involve inflammation response in biotype 2, affecting fundamental brain functions by compromising synaptic plasticity, neurogenesis, neuronal differentiation, and migration.

The core pathology of biotype 3 may be the impaired neuromodulation of synaptic signaling transmission, leading to abnormal functional integration of neural systems. Interestingly, besides the risk genes related to the transport of calcium ions and neurotransmitters, such as *PTPRD* and *CACNA2D1*, we identified *CDKAL1*, a gene susceptible to type 2 diabetes. The gene hosts 38.8% SNP alleles in the top-ranking subtype-specific PHB with the highest OR, indicating that *CDKAL1* may be a shared genetic risk factor for both diseases. Additionally, biotype 3 had different neuroimaging features, ReHo was significantly increased in the frontal lobe and significantly decreased in the parietal lobe, postcentral gyrus, precentral gyrus, and temporal lobe compared to HCs.

**MATERIALS AND METHODS**

*Subjects and data collection and preprocessing*

A cohort of 141 individuals with SCZ diagnosis was recruited from the Shenyang Mental Health Center and the First Affiliated Hospital of China Medical University in Shenyang, China. In addition, 284 healthy controls (HCs) were recruited from local communities. Behavioral symptoms were assessed using the Brief Psychiatric Rating Scale (BPRS). Cognitive function was assessed using the Wisconsin Card Sorting Test (WCST) and MATRICS Consensus Cognitive Battery (MCCB). Demographic and clinical characteristics are detailed in Supplementary eTable 1. Two professionally trained psychiatrists carried out all diagnoses following the structured clinical interview for DSM-IV Axis I Disorders for participants over 18 and the Schedule for Affective Disorders and Schizophrenia for School-Age Children-present and Lifetime Version (K-SADS-PL) for participants under 18 years old. All patients met DSM-IV criteria for SCZ and had no other comorbid Axis I disorders. The HCs had no personal history of psychotic or recurrent mood disorder. Exclusion criteria for the study included (1) the presence of organic brain disease, (2) a history of substance abuse or dependence (except tobacco), (3) a history of head injury, epilepsy, or coma, and (4) contraindications for magnetic resonance examination. The Institutional Review Board of China Medical University approved the study, and informed consent was obtained from all participants.

*Genotyping and Quality Control*

Peripheral venous blood samples (5 mL) were collected from each subject using EDTA Vacutainer tubes and centrifuged at 2000 rpm for 10 min. Subsequently, plasma and blood cells were separated and stored at -80°C for genomic (100 μL each) analyses. All participants were genotyped using whole blood and the Illumina Global Screening Array-24 v1.0 BeadChip, covering 642,824 fixed genetic variants and 53,411 customized variants specific to the Han Chinese populations. The assays were performed following the manufacturer's recommendations, and PLINK v1.9[35] was employed for quality control of the genotyping data. Single nucleotide polymorphisms (SNPs) with minor allele frequency (MAF) less than 1%, call rate less than 95% or Hardy-Weinberg equilibrium (HWE) *p*-value less than $10^{-5}$ were excluded. Individuals with an excessive rate of missing data, more significant than 5%, were removed. The final genotyping data included 404,078 genetic variants for 424 participants.

*Functional MRI Acquisition and Preprocessing*

Magnetic Resonance Imaging (MRI) scans were obtained using a GE Signa HD 3.0T scanner with a standard 8-channel head coil at the First Affiliated Hospital of China Medical University in Shenyang, China. The functional images were acquired using a spin echo planar imaging (EPI) sequence, with the following parameters: repetition time (RT) = 2,000 ms, echo time (TE) = 30 ms, flip angle = 90°, the field of view = 240×240 mm$^2$, matrix = 64 x 64. A total of 35 axial slices were collected with 3 mm thicknesses without a gap.

The Functional MRI (fMRI) preprocessing was performed using the Statistical Parametric Mapping 8 (SPM8; http://www.fil.ion.ucl.ac.uk/spm) and the Data Processing Assistant for R-functional MRI (DPARSF; http://www.restfmri.net/forum/DPARSF)[6,63]. The first 10 time points of functional images were discarded to ensure magnetization stabilization, followed by slice timing correction, motion correction, and spatial normalization. The images were then normalized to Montreal Neurological Institute (MNI) space using a standard EPI template with a re-sampling voxel volume of 3 mm x 3 mm x 3 mm. The images were then spatially smoothed with a 6 mm full-width-at-half-maximum (FWHM) Gaussian kernel. To minimize the effects of frequency drifts and high-frequency noise, all-time courses were filtered between 0.01–0.08 Hz. After preprocessing, we applied DPARSF to calculate Regional Homogeneity (ReHo), which was used to measure the similarity of the time series of a given voxel to those of its adjacent 26 voxels. The whole brain average normalized the ReHo values.

**Identifying disease subtypes based on polygenic modules**

The Custom Correlation Coefficient (CCC)[27] was adopted to construct an SNP-allelic network for a given genotyping dataset. In the allelic network, an SNP is represented by two nodes, one for each SNP allele. An edge is introduced to link two alleles of different SNPs if their correlation is above a significance threshold. In our current implementation, we applied a stringent CCC threshold of 0.68 to construct the network using data of all patients for subtyping. The Louvain community detection algorithm[31] was applied to the network to identify SNP-allele modules with tight connections within a module and relatively sparse connections across modules. Louvain is a multi-phase and iterative heuristic module detection algorithm for optimizing the Q-function of global modularity, which describes the tightness of the modules[31,64]. Compared with other community methods, Louvain can extract modules in large networks with less computation.

To identify candidate polygenic biomarkers for SCZ, we computed the frequencies, odds ratios (ORs), and risk ratios (RRs) for every module for SCZ subjects and HCs. The OR was calculated as OR = a(d-b)/b(c-a), where a and b denote the number of SCZ cases and HCs, respectively, who hold the module of interest, and c and d represent the total number of cases and HCs who possess complete data for the SNPs in the module, respectively. The RR was computed as RR = a/b. The 95% confidence interval (CI) was defined as $e$^[$ln$(OR)±1.96$sqrt$(1/a+1/b+1/(c-a)+1/(d-b))], where $ln$ is the natural logarithm and $sqrt$ is the square root. We selected those SNP-allele modules with OR ⩾ 1.2 and RR ⩾ 1.2 as risk SNP-allele modules.

### *Identify subtype-specific polygenic haplotype biomarkers*

We used a more stringent CCC threshold of 0.72 to build an SNP-allele network for every biotype to identify SNP-allele modules. We then adopted a set of statistical tests to search for statistically significant SNP-allele modules as subtype-specific polygenic haplotype biomarkers (PHBs).

*Permutation trials for avoiding false-positive edges*

To avoid false-positive edges in an SNP-allele network, we randomly shuffled the genotypes of the individuals with the original alleles and genotype frequency of every SNP intact. After randomization, we recalculated the CCC value for every pair of SNP alleles and filtered out the false-positive edge if the CCC value for one pair of SNP alleles was smaller than the CCC threshold in the original network after multiple permutations.

*Permutation trials*

We randomly permuted the phenotype labels across the subtype of SCZ and HCs. We then computed the G-test score by G-test of independence[65] for every module for these randomized groups of individuals and repeated these trials 1,000 times. We retained the modules with statistically significant $p$-values (<0.05) for the G-test scores and discarded the modules that failed the permutation test.

*Bootstrapping trials*

We randomly selected eighty percent of the cases and eighty percent of the HCs, and used these samples to compute the ORs for the modules of every subtype. We repeated these trials 1,000 times. We calculated the OR and $p$-value for every module in each trial and computed the mean OR and average 95% confidence intervals over the 1,000 trials for every module in each subtype. We remove the modules with ORs <1.5 or 95% confidence interval ≤ 1.05 over the 1,000 random samplings.

We looked for subtype-specific PHBs following the same scheme as finding SNP-allele modules for SCZ. Instead of using all SCZ cases, only those belonging to a biotype were used to construct a new, subtype-specific SNP-allele network. We also used more stringent criteria to select subtype-specific PHBs: OR ≥ 1.5, a lower limit of 95% CI ≥ 1.1, and Yates's correction $p$-value < 0.05. Additionally, all the subtype-specific PHBs must pass the permutation and bootstrapping tests.

### *Subtype validation using multimodal biological data*

To assess the diagnostic capability of PHB-based subtyping, we compared PHBs before and after subtyping using additional data and information from multiple sources.

*Gene function enrichment analysis*

A gene set enrichment analysis was performed using the web-based platform FUMAGWAS[34] on the genes hosting the SNP alleles in a PHB or genes closest to the SNPs. We extended SNPs in noncoding transcripts' genomic coordinates upstream 15kb and downstream 10kb to include adjacent genes. The enrichment analysis utilized molecular pathways from the biological processes of gene ontology (GO)[66], and biological Processes (BP) with FDR-corrected *p*-value < 0.05 were regarded as significantly enriched.

*Comparison of ReHo alterations between subtypes and HC*

To identify ReHo alteration in fMRI across subtypes and HCs, we used the brain imaging data processing and analysis tool DPABI[67] to conduct a two-sample t-test to compare the ReHo values between each subtype and HCs. Gender and age were covariates in the two-sample t-test with statistical significance of *p*-values < 0.05 after the Gaussian random field (GRF) correction.

*Comparison of clinical and cognitive characteristics of subtypes*

One-way ANOVA analysis was adopted to determine if significant differences existed between subtypes and HCs in cognitive ability measured by the Wisconsin Card Sorting Test (WCST) and MATRICS Consensus Cognitive Battery (MCCB) and in behavioral symptoms assessed by Brief Psychiatric Rating Scale (BPRS). The statistical significance level was set at *p*-values < 0.05 after the Least Significant Difference (LSD) correction. The WCST and MCCB are routine neurocognitive tasks frequently used to assess cognitive flexibility in patients with mental disorders[68-70], especially SCZ. Five WCST performance measures were used to access cognitive abilities: the numbers of categories completed (CC), correct responses (CR), total errors (TE), perseverative errors (PE, i.e., the subject persisted in making an incorrect sorting choice), and non-perseverative errors (NPE, i.e., incorrect responses other than PE). Higher scores of CC and CR indicated good cognitive performance, while higher scores of TE, PE and NPE suggested poor cognitive function. MCCB includes seven psychological dimensions and ten subtests. All test scores were converted to T-scores (T1-T10, mean 50, standard deviation 10) based on original scores. A higher T-score for MCCB indicates good cognitive function.

Brief Psychiatric Rating Scale (BPRS) factor scores were identified from the exploratory factor analysis (EFA) using the principal component analysis method[32], which resulted in a parsimonious list of factors using BPRS items. Five interpretable and clinically relevant factors were identified: Anxiety and Depression, Negative symptoms, Hostility-suspicion, Activation, and Thinking Disorder[71].

*Competing interests*

The authors declare that they have no competing interests

*Funding*

This work was supported by the National Natural Science Foundation of China (grant 62176129), the National Key Research and Development Program of the Ministry of Science and Technology of China (2022YFC2405603), National Science Fund for Distinguished Young Scholars (81725005), NSFC-Guangdong Joint Fund (U20A6005), Jiangsu Provincial Key Research and Development Program of Science and Technology Department of Jiangsu Province (BE2021617), the National Key R&D Program of China (2018YFC1311600 and 2016YFC1306900) and Liaoning Revitalization Talents Program (XLYC1808036). The work was also supported by the Hong Kong Health and Medical Research Fund (HMRF grant 10211696), the Hong Kong Global STEM Professor Scheme, and the Hong Kong Jockey Club Charities Trust.

*Author contributions*

Xizhe Zhang, Fei Wang, and Weixiong Zhang conceived and designed the research. Xinru Wei implemented the algorithm, collected and analyzed the data, and helped write the manuscript and prepare the figures. Lili Tang, Zhao Su, Pengfei Zhao helped prepare the figures. Yanqing Tang, Shuai Dong, Chunyu Pan collected the data. Xizhe Zhang and Weixiong Zhang analyzed the data and drafted the manuscript.

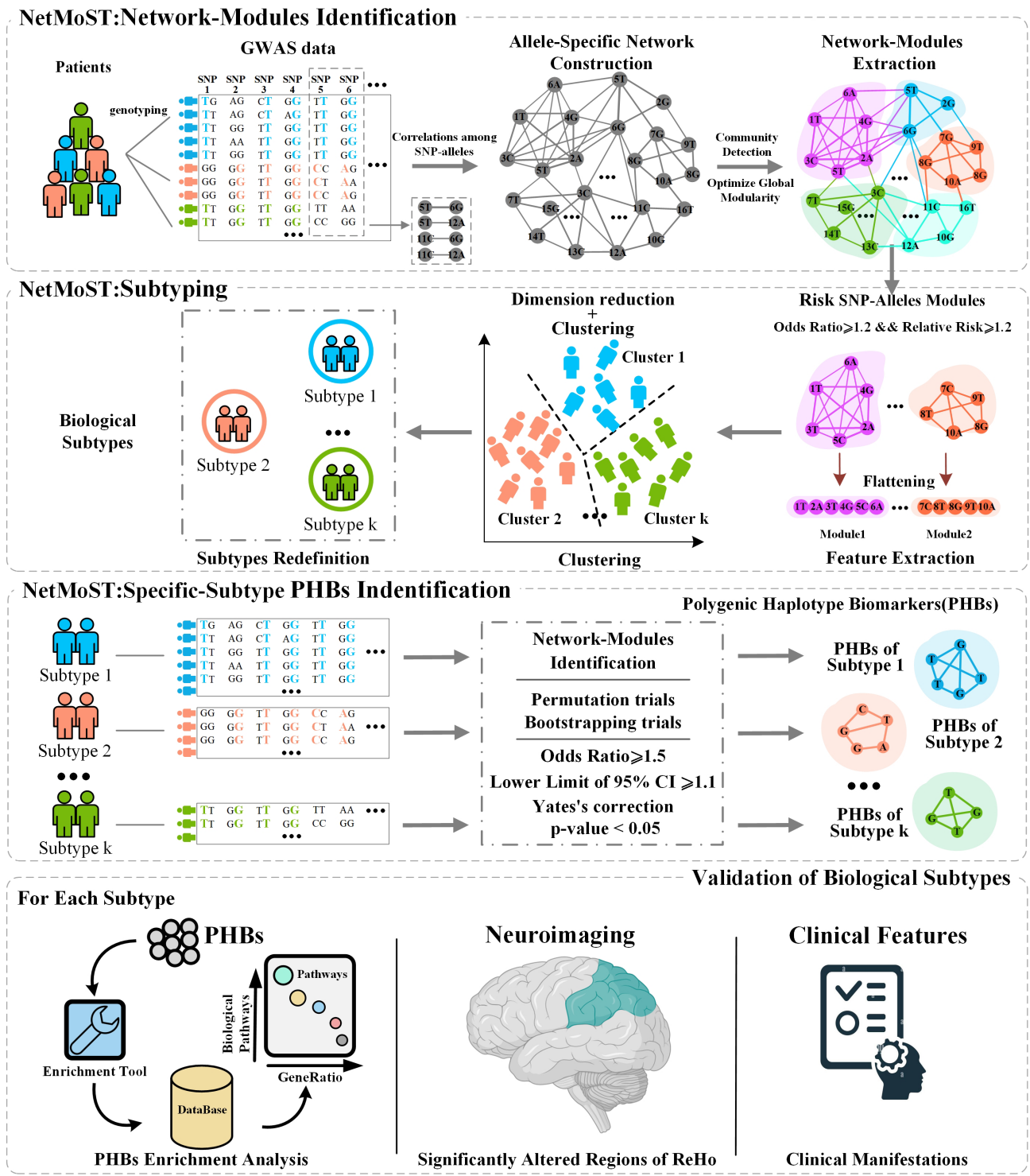

**Figure 1 | Schematic of the NetMoST framework to define and validate objective subtypes of complex diseases.** Stage one: Construction of specific SNP-allele network for complex diseases like SCZ and using Louvain community detection to extract risk SNP-allele modules. Stage two: For the risk SNP-allele modules of schizophrenia from stage one, the clustering method was applied to discover subtypes of SCZ. Stage three: The genetic variants of cases belonging to each subtype were used to construct a new, subtype-specific SNP-allele network. Then the more stringent criteria (OR $\geq$ 1.5, the lower limit of 95% CI $\geq$ 1.1, and Yates's correction p-value < 0.05) to select subtype-specific PHBs that also underwent permutation and bootstrapping trials. Finally, the biological subtypes are validated utilizing multimodal data, including genetic profiles, brain neuroimaging, and clinical features. Firstly, the gene set enrichment analysis was applied to these PHBs for each subtype to

discover the unique genetic pattern. Furthermore, the neuroimaging and clinical features were examined to determine other biological distinctions within the allele-module-based subtypes.

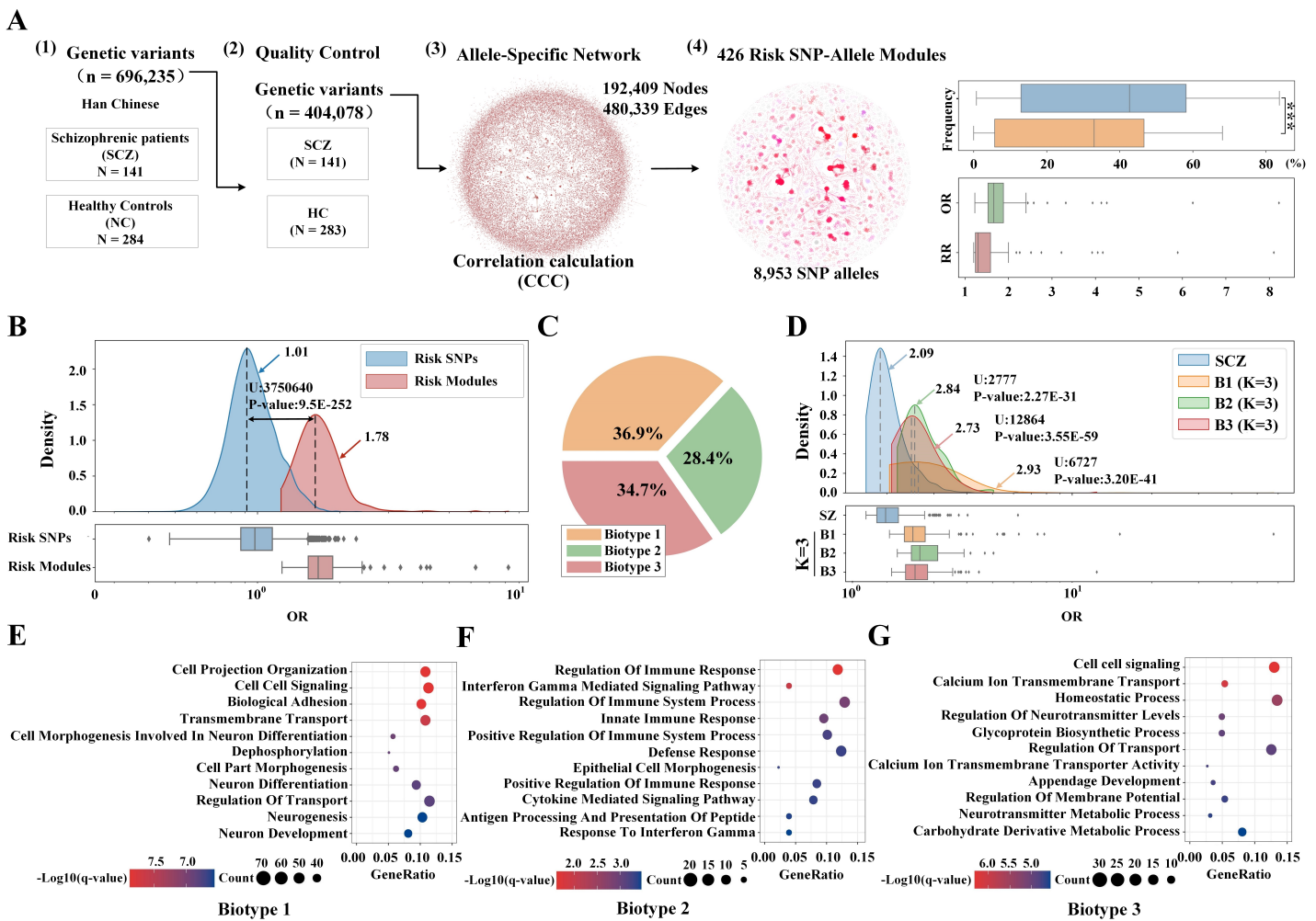

**Figure 2 | NetMoST method was applied to subtype schizophrenia (SCZ) and define three biological subtypes of SCZ based on risk SNP-alleles modules. A)** standard quality control procedures were performed on the genetic variations dataset. Finally, 404,078 SNPs were retained, and 141 and 283 subjects were labeled as schizophrenic cases and healthy controls, respectively. Based on these SNP alleles that passed quality control, a specific SNP-allele network with 192,409 nodes (SNP-alleles) and 480,339 edges (associations of SNP-allele pairs) was established for SCZ, and 426 risk SNP-alleles modules (OR ≥1.2 and RR≥1.2) highly associated with SCZ were identified. **B)** These single SNPs within the risk modules had lower OR values than the OR of risk SNP-allele modules, which indicates the necessity of capturing combinatorial genetic risk for complex disease. **C)** The quantitative distribution of three biotypes for SCZ subjects. The netMoST framework identified three biological subtypes, with biotypes 1, 2, and 3 having 52, 40 and 49, respectively. **D)** The odds ratios of PHBs of each subtype after subtyping were significantly higher than before subtyping. This further explained that accurate classification would improve the power to recognize polygenic signatures. **E-G)** Significant biological characteristics across three biotypes of SCZ. Examining biological characteristics from genomics between each biotype and HCs revealed that the three biotypes had unique genetic patterns. Gene set enrichment of PHBs for each biotype. The PHBs in biotype 1 are involved in nervous system development, including neuron development, differentiation, and neurogenesis. The PHBs in biotype 2 are significantly enriched in the process of related-immune function, and the PHBs in biotype 3 appear in the transports of calcium ions and neurotransmitters.

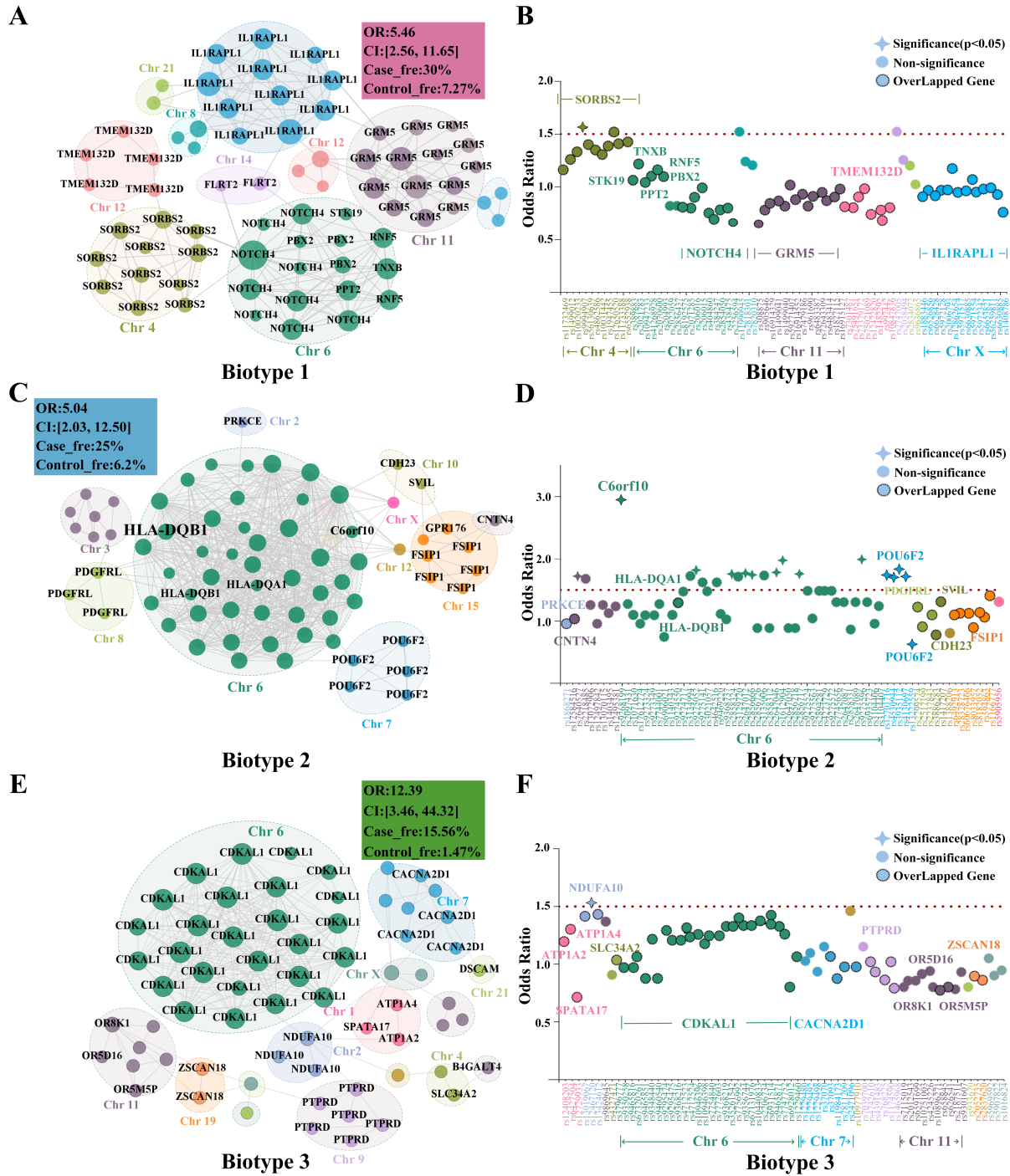

**Figure 3 | The subtype-specific polygenic haplotype biomarker capable of exhibiting unique genetic characteristics for each biotype of SCZ.** The novel netMoST method was introduced to discover genetic subtypes based on SNP-alleles modules and identified three biotypes of SCZ. The subtype-specific representative PHB with higher OR capable of exhibiting unique genetic characteristics for each biotype was extracted by constructing specific SNP-alleles networks and community detection. **A, C, E)** Topological network structure for SNP alleles (Each node label was named by its overlapping gene) of subtype-specific PHB with higher OR for each biotype. Each SNP allele is represented by a node, where its size is proportional to the node degree in this network, and its color represents its chromosome. **B, D, F)** The odds ratio of each SNP of the subtype-specific PHBs with higher OR of each biotype. The horizontal axis represents the SNP for SPHB of each subtype, and the vertical axis represents the odds ratio of SNP. Each SNP is represented by a node whose color represents its chromosome. The three representative PHBs reveal that netMoST could identify combinatorial genetic risk factors easily

neglected by conventional single-marker-based methods due to the smaller OR. These genetic risk factors may significantly form significant haplotypes with long-range linkage associations by spanning multiple chromosomes.

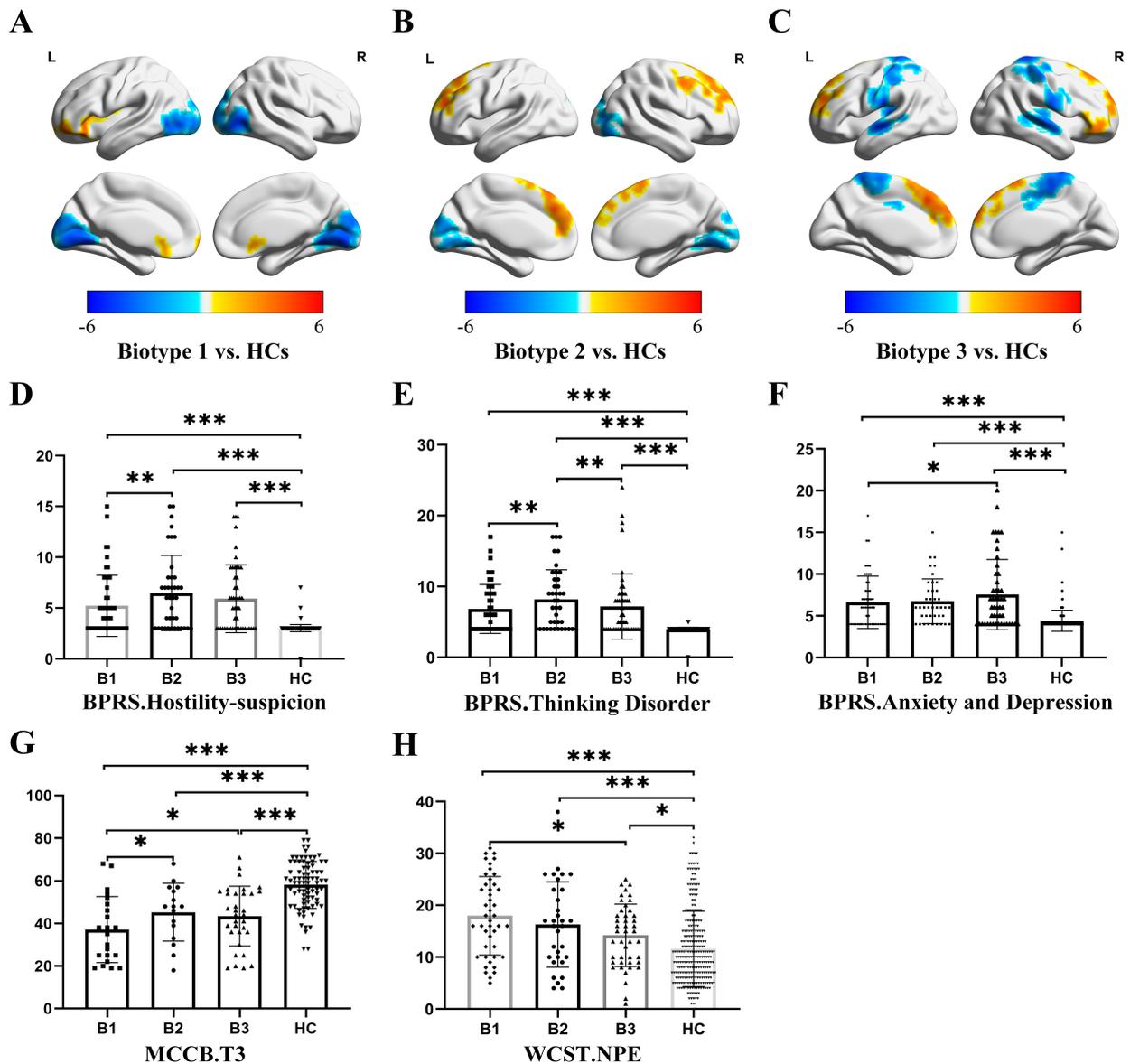

**Figure 4 | Significant differences in brain neuroimaging and clinical characteristics among three biotypes of SCZ. A-C)** The neuroimaging regional homogeneity (ReHo) alterations between each biotype and HCs. **D-F)** The differences in behavioral symptoms assessed by the Brief Psychiatric Rating Scale (BPRS) among three biotypes and HCs. **G-H)** The cognitive ability between three biotypes of SCZ. Biotype 1 performed worse on cognitive measures than the other two biotypes due to the lower score of T3 and the greater score of NPE belonging to the MATRICS Consensus Cognitive Battery (MCCB) and the Wisconsin Card Sorting Test (WCST), respectively. The vertical black lines show the standard errors of the means, and the significance level was set to $p < 0.05$ with the Least Significant Difference correction. ***$p < 0.001$; **$p < 0.01$; *$p < 0.05$.